\documentclass{kapproc} 

\newread\epsffilein    
\newif\ifepsffileok    
\newif\ifepsfbbfound   
\newif\ifepsfverbose   
\newdimen\epsfxsize    
\newdimen\epsfysize    
\newdimen\epsftsize    
\newdimen\epsfrsize    
\newdimen\epsftmp      
\newdimen\pspoints     
\def\epsfbox#1{\global\def\epsfllx{72}\global\def\epsflly{72}%
   \global\def\epsfurx{540}\global\def\epsfury{720}%
   \def\lbracket{[}\def\testit{#1}\ifx\testit\lbracket
   \let\next=\epsfgetlitbb\else\let\next=\epsfnormal\fi\next{#1}}%
\def\epsfgetlitbb#1#2 #3 #4 #5]#6{\epsfgrab #2 #3 #4 #5 .\\%
   \epsfsetgraph{#6}}%
\def\epsfnormal#1{\epsfgetbb{#1}\epsfsetgraph{#1}}%
\def\epsfgetbb#1{%
\openin\epsffilein=#1
\ifeof\epsffilein\errmessage{I couldn't open #1, will ignore it}\else
   {\epsffileoktrue \chardef\other=12
    \def\do##1{\catcode`##1=\other}\dospecials \catcode`\ =10
    \loop
       \read\epsffilein to \epsffileline
       \ifeof\epsffilein\epsffileokfalse\else
          \expandafter\epsfaux\epsffileline:. \\%
       \fi
   \ifepsffileok\repeat
   \ifepsfbbfound\else
    \ifepsfverbose\message{No bounding box comment in #1; using
defaults}\fi\fi
   }\closein\epsffilein\fi}%
\def\epsfsetgraph#1{%
   \epsfrsize=\epsfury\pspoints
   \advance\epsfrsize by-\epsflly\pspoints
   \epsftsize=\epsfurx\pspoints
   \advance\epsftsize by-\epsfllx\pspoints
   \epsfxsize\epsfsize\epsftsize\epsfrsize
   \ifnum\epsfxsize=0 \ifnum\epsfysize=0
      \epsfxsize=\epsftsize \epsfysize=\epsfrsize
arithmetic!
     \else\epsftmp=\epsftsize \divide\epsftmp\epsfrsize
       \epsfxsize=\epsfysize \multiply\epsfxsize\epsftmp
       \multiply\epsftmp\epsfrsize \advance\epsftsize-\epsftmp
       \epsftmp=\epsfysize
       \loop \advance\epsftsize\epsftsize \divide\epsftmp 2
       \ifnum\epsftmp>0
          \ifnum\epsftsize<\epsfrsize\else
             \advance\epsftsize-\epsfrsize \advance\epsfxsize\epsftmp \fi
       \repeat
     \fi
   \else\epsftmp=\epsfrsize \divide\epsftmp\epsftsize
     \epsfysize=\epsfxsize \multiply\epsfysize\epsftmp
     \multiply\epsftmp\epsftsize \advance\epsfrsize-\epsftmp
     \epsftmp=\epsfxsize
     \loop \advance\epsfrsize\epsfrsize \divide\epsftmp 2
     \ifnum\epsftmp>0
        \ifnum\epsfrsize<\epsftsize\else
           \advance\epsfrsize-\epsftsize \advance\epsfysize\epsftmp \fi
     \repeat
   \fi
   \ifepsfverbose\message{#1: width=\the\epsfxsize,
height=\the\epsfysize}\fi
   \epsftmp=10\epsfxsize \divide\epsftmp\pspoints
   \vbox to\epsfysize{\vfil\hbox to\epsfxsize{%
      \includegraphics{#1}%
      \hfil}}%
\epsfxsize=0pt\epsfysize=0pt}%

{\catcode`\%=12
\global\let\epsfpercent=
\long\def\epsfaux#1#2:#3\\{\ifx#1\epsfpercent
   \def\testit{#2}\ifx\testit\epsfbblit
      \epsfgrab #3 . . . \\%
      \epsffileokfalse
      \global\epsfbbfoundtrue
   \fi\else\ifx#1\par\else\epsffileokfalse\fi\fi}%
\def\epsfgrab #1 #2 #3 #4 #5\\{%
   \global\def\epsfllx{#1}\ifx\epsfllx\empty
      \epsfgrab #2 #3 #4 #5 .\\\else
   \global\def\epsflly{#2}%
   \global\def\epsfurx{#3}\global\def\epsfury{#4}\fi}%
\def\epsfsize#1#2{\epsfxsize}
\def\eps@scaling{.95}
\def\epsscale#1{\gdef\eps@scaling{#1}}
\def\plotone#1{{\centering \leavevmode
    \epsfxsize=\eps@scaling\columnwidth \hfil \hbox{\epsfbox{#1}}\hfil}}

\def\ts{\thinspace}

\let\footnote\savefootnote

\setcounter{tocdepth}{1}

\kluwerbib

\begin{document}



\articletitle[ULIGs and the Origin of 
QSOs]{Ultraluminous Infrared Galaxies and the Origin of QSOs} 

\chaptitlerunninghead{ULIGs and the Origin of QSOs}

\author{D. B. Sanders$^{1,2}$}
\affil{$^1$Institute for Astronomy, Univ. of Hawaii, 2680 Woodlawn 
Drive, Honolulu, HI  96822}
\affil{$^2$Max-Plank Institut f$\ddot u$r Extraterrestrische Physik, 
D-85740 Garching, Germany}
\email{sanders@ifa.hawaii.edu}



\begin{abstract}
We review the evidence which suggests that ultraluminous infrared galaxies
(ULIGs) are the precursors of optically selected quasi-stellar objects
(QSOs) and discuss additional data that suggests that the majority, if not
all QSOs, may begin their lives in an intense infrared phase.
Implications for the host galaxies of QSOs are discussed.
\end{abstract}


\section{Introduction}
The discovery of a significant population of ultraluminous infrared
galaxies (ULIGs) whose infrared luminosities, $L_{\rm ir} \equiv
L(8-1000\mu m) > 10^{12} L_\odot$, are equivalent to the bolometric
luminosities of optically selected QSOs, [i.e.  $M_{\rm B} < -22.12$,
$(H_{\rm o},q_{\rm o}){\ts}=$(75,0), equivalent to $M_{\rm B} < -23.0$,
(Schmidt \& Green 1983)], was one of the major scientific results from the
Infrared Astronomical Satellite ({\it IRAS}) all-sky survey (Soifer et al.
1987).  Extensive ground-based follow-up studies of complete samples of
the nearest ULIGs have clearly shown that strong interactions/mergers of
molecular gas-rich spirals are responsible for triggering the intense
infrared emission, which appears to be produced by a mixture of nuclear
starburst and powerful AGN activity.  It has been postulated that the AGN
may dominate the $L_{\rm ir}$ output in the ULIG phase and that the
majority of ULIGs may evolve into QSOs (e.g. Sanders et al. 1988a).

Here we briefly summarise the most recent data for complete samples of
nearby ULIGs in order to illustrate the major stages in the proposed
evolutionary scenario from ULIG to QSO. New data on the properties of the
host galaxies of ULIGs are presented and the consequences for the host
galaxies of QSOs are discussed.

\section{An Evolutionary Scenario:\ ``cool" ULIGs $\to$ ``warm" 
ULIGs $\to$ ``IR-excess" QSOs}

\begin{figure}
\includegraphics{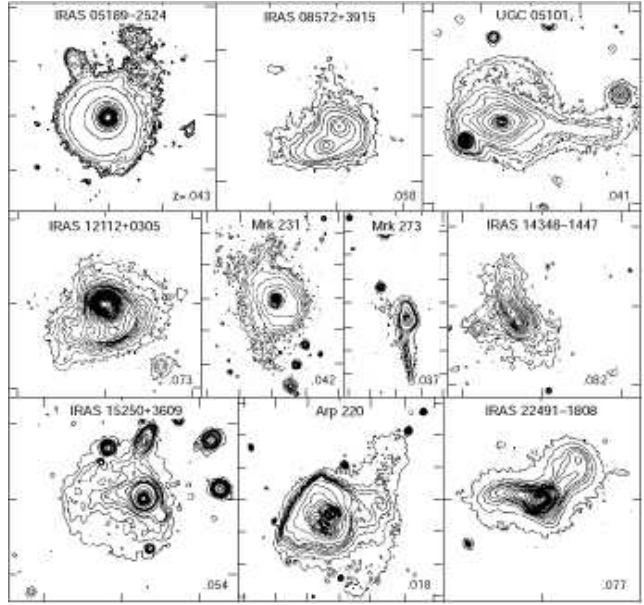}
\vspace{8.0cm}
\caption{Optical (r-band) CCD images of the complete sample of 10 ULIGs
from the original BGS (Sanders et al. 1988a).  Tick marks are at
20$^{\prime\prime}$ intervals.
}
\end{figure}

What are the range of properties exhibited by ULIGs?  Where did they come
from and what are they evolving into ?  Extensive multiwavelength studies
of the complete sample of objects in the IRAS BGS have provided the most
detailed answers to these questions concerning the origin and evolution of
ULIGs and their relationship to other classes of extragalactic objects.
Sanders \& Mirabel (1996) provide a comprehensive review of the available
data on ULIGs, but for our purposes here Figures 1-5 will be sufficient to
describe the general properties of ULIGs and to introduce our proposed
evolutionary scenario.

The images shown in Figure 1 illustrate the range of morphology seen in
ULIGs.  Analysis of this sample of 10 ULIGs from the original IRAS BGS
shows that all 10 are strongly interacting/merger pairs involving spiral
disks of relatively equal mass ($<$3:1) near the endstage of the merger
when the two disks substantially overlap and the two nuclei are close to
or have already merged.  Studies of larger samples of ULIGs have reported
some variations on this theme, notably ULIGs with widely separated
non-overlapping disks (e.g. Murphy et al. 2001) and ULIGs composed of
three or more interacting disks (e.g. Borne et al. 2000), both phenomena
being observed in $\sim$15-20\% of the objects studied; however our own
extensive study of a sample of 118 ULIGs suggests that such variations, if
true, apply to a much smaller fraction ($<$3-5\%) of the total population
of ULIGs in the local universe (Veilleux, these proceedings).

\begin{figure}
\includegraphics{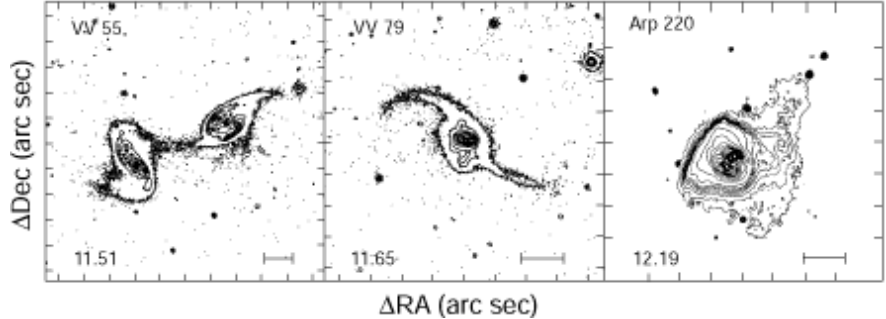}
\vspace{5.0cm}
\caption{R-band images of LIGs (Mazzarella et al. 2001) from the IRAS BGS
illustrating the strong interactions/mergers that are characteristic of
nearly all objects with $L_{\rm ir} > 10^{11.3} L_\odot$.  The scale bar
represents 10{\ts}kpc, tick marks are at 20$^{\prime\prime}$ intervals,
and the value of $L_{\rm ir}$ is indicated in the lower left corner of
each panel.
}
\end{figure}

Objects in the pre-ULIG phase can be seen in the images of the complete
IRAS BGS.  Most objects with $L_{\rm ir} 10^{11.3}{\ts}L_\odot$ appear to
fall into a merger sequence which may result in a ULIG phase; $L_{\rm ir}$
generally increases with decreasing projected nuclear separation (see
Figure 2).  All of these merger systems involve disks which are both
molecular gas rich.  The fate of the molecular gas appears to be similar
to what has been predicted from numerical simulations (e.g. Barnes \&
Hernquist 1991); enhanced star formation is observed as the gas disks
collide, but eventually $\sim$40-70\% (depending on encounter geometry,
relative velocity, etc.) of the total initial molecular gas mass is
funnelled into a dense nuclear core (typically $R < 1$kpc).  Figure 3
gives a dramatic illustration of such an end product, with remnant star
clusters still visible in the larger circumnuclear environs ($R < 5$kpc)
and a central dense gas core containing more than $10^9\ M_\odot$ of H$_2$
gas.  The bulk of the total $L_{\rm ir}$ in Mrk231 originates within the
central $R < 0.5${\ts}kpc (Soifer et al. 2000).  Whether a powerful
nuclear starburst or accretion onto a massive black hole (MBH) dominates
within the dense nuclear gas core is a subject of current debate (c.f.
Joseph 1999; Sanders 1999), but it is clear that the conditions are
favourable for both.

\begin{figure}
\vspace{-0.5cm}
\includegraphics{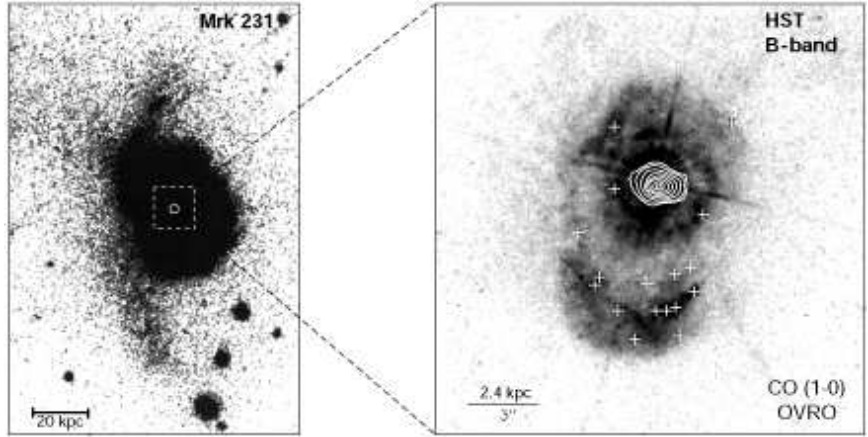}
\vspace{5.5cm}
\caption{The ``IR-QSO" Mrk{\ts}231.  (left panel) - optical image and size
of central molecular gas concentration (circle).  (right panel) - HST
B-band image and identified stellar clusters (`+') from Surace et al.
(1998).  The high resolution CO contours are from Bryant \& Scoville
(1996).
}
\end{figure}

\begin{figure}
\vspace{1.0cm}
\includegraphics{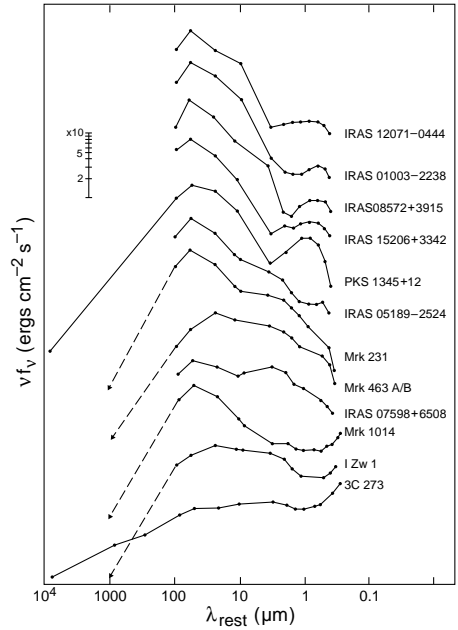}
\vspace{7cm}
\caption{SEDs for the complete sample of 12 ``warm" ULIGs from Sanders et
al. (1988b).  Sources are ordered (top to bottom) in order of increasing
$f_{25}/f_{60}$ ratio and increasing ratio of $L_{\rm opt-UV} / L_{\rm
ir}$.  Dashed lines are an extrapolation beyond the figure boundary to the
data point at $\lambda = 6${\ts}cm.
}
\end{figure}

A significant subset of ULIGs, those exhibiting ``warm" mid-infrared
colors (i.e. $f_{25} / f_{60} > 0.2$)\footnote{deGrijp et al. (1985) found
that searches based on ``warm" mid-infrared colors could be very useful
for discovering new infrared-luminous Seyfert galaxies in the IRAS
database} give the strongest evidence favouring the hypothesis that the
dominant luminosity source in the ULIG phase is an AGN.  The three ``warm"
ULIGs in Figure 1 (IRAS05189-2524, IRAS08579+3925, Mrk231) all show broad
Sy{\ts}1 emission lines in optical and/or near-infrared spectra (Veilleux
et al. 1995), and the ratio of reddening-corrected H$\beta$ broad-line
luminosity to $L_{\rm bol}$ is the same in these objects as is found for
optically selected QSOs (Veilleux et al. 1999).  In larger samples, the
ratio of ``cool" to ``warm" ULIGs increases with increasing $L_{\rm ir}$,
from $\sim$15\% at $10^{12}{\ts}L_\odot$ to $\sim$50\% at $>
10^{12.4}{\ts}L_\odot$ (see Veilleux, these proceedings).  The majority of
the ``warm" objects ($\sim$75\%) exhibit a single merger nucleus as
opposed to the ``cool" objects where the mean nuclear separation is
$\sim$2kpc and only $\sim$20\% of the sources exhibit a single nucleus
(e.g. Kim 1995).

\begin{figure}
\vspace{-0.5cm}
\includegraphics{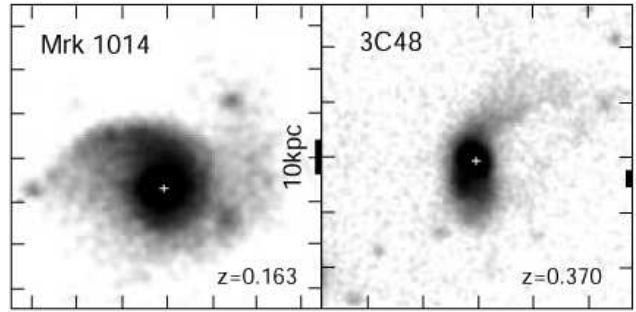}
\vspace{5cm}
\caption{Optical images of ``IR-excess", optically selected QSOs
(Mrk{\ts}1014 ={\ts}PG0157+001 - MacKenty \& Stockton 1984; 3C48 -
Stockton \& Ridgway 1991)
} 
\end{figure}

As in the small ``warm" subsample of 3 ULIGs shown in Figure 1, larger
samples of ``warm" ULIGs show broad Sy{\ts}1 emission lines either in
direct light or in polarised emission.  And more importantly, a
significant fraction ($\sim$10--15\%) already exhibit the ``big-blue-bump"
characteristic of optically-selected QSOs.  Figure 4 shows the complete
radio-to-UV spectral energy distributions (SEDs) for a complete sample of
12 ``warm" ULIGs (Sanders et al. 1988b) illustrating the  progression from
strong far-infrared excess through equally strong mid-infrared emission
and finally to the emergence of a UV-excess phase.  It seems reasonable to
assume that the emergence of the ``big-blue-bump" may be associated with
the expulsion of gas and dust from the central regions through the
combined action of radiation pressure and supernova explosions as
indicated by the detection of nuclear superwinds (e.g. Armus et al. 1990)
in many of these objects.  Figure 5 shows images of two ``warm" ULIGs
which are also well-known QSOs (one RQQ and one RLQ).  Both of these
objects still show remnant tidal tails and both still contain relatively
large amounts of molecular gas, consistent with the finding by Clements
(2000) that those QSOs with the strongest far-infrared excesses are more
likely to lie in visibly ``disturbed" hosts.

\section{The Host Galaxies of ULIGs: Evidence for $S + S \to E$}

\begin{figure}
\vskip 0.5truecm
\includegraphics{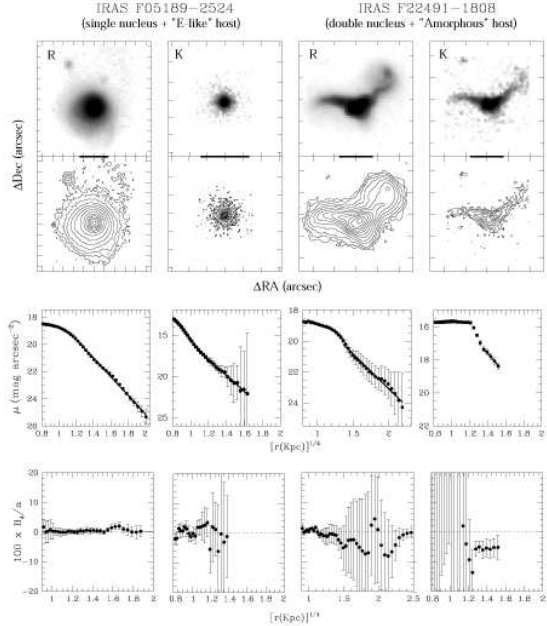}
\vspace{7.5cm}
\caption{R-band and K-band images and 1-D surface intensity profiles for
two ULIGs - the ``IR-QSO" IRAS{\ts}F05189$-$2524 and the ``cool" ULIG
IRAS{\ts}22491$-$1808 (Sanders et al. 2000)
} 
\end{figure}

High resolution optical and near-infrared images of the hosts of ULIGs
from the IRAS BGS have shown that they appear to be evolving into
elliptical galaxies with optical half-light radii and surface brightness
values which lie on the fundamental plane defined by ellipticals (Kim
1995).  Furthermore, the central gas surface densities of ULIGs are
similar to the values found in disky ellipticals, thus solving a long
standing problem of how such dense cores could have formed (e.g. Kormendy
\& Sanders 1992).  And Genzel et al. (2001) ``confirm that ULIG mergers
are ellipticals-in-formation" based on near-infrared spectroscopy of 12
ULIGs from the IRAS BGS, which shows that these objects fall on or near
the fundamental plane represented by intermediate mass ($\sim{\ts}L^*$)
disky ellipticals.

A more thorough analysis of the radial surface brightness distributions
for all 118 ULIGs in the IRAS 1-Jy sample suggests that roughly 30\% of
the R-band radial surface brightness profiles are well fitted by an
elliptical-like $R^{1/4}$-law, with only 10\% showing an exponential
profile and 60\% being ``intermediate" (see Veilleux, these proceedings).
Figure 6 gives examples of the observed radial profiles for a ``warm"
``E-like" ULIG and a ``cool" ``intermediate-type" ULIG.  The ``E-like"
ULIGs still show evidence of dusty nuclei as evidenced by flat-topped
R-band profiles and excess K-band light at $R <${\ts}1--2{\ts}kpc, and the
outermost radii are often affected by faint tidal debris.
``Intermediate-type" ULIGs (the majority of all ULIGs) often show the
effects of double nuclei at small radii ($R < 5$kpc), a $R^{1/4}$-law over
much of the rest of the observable disk, but with strong asymmetries due
to obvious tidal debris at large radii.

\section{Comparison of ULIG/QSO Host Magnitudes and Colors}

There is now sufficient high resolution imaging data on large samples of
QSOs and ULIGs to permit a meaningful comparison of the basic properties
of the hosts of both classes of objects, although caution is advised
depending on which samples are selected.  There is still a mismatch in
redshift and bolometric luminosity distribution between the two classes of
objects depending on which published samples are chosen. Problems with PSF
subtraction in QSO images limits much of the analysis of QSO data to 1-D
radial profiles and results in host magnitudes that suffer from somewhat
uncertain extrapolations to small radii.

H-band imaging is thought to minimise the contamination from QSO nuclei so
we present these results first.  For QSOs, extensive H-band imaging of
local (typically $z < 0.3$) ``low- luminosity" QSOs shows that they
typically lie in $\sim${\ts}$L^*$ host galaxies (Mcleod \& Rieke 1994a)
while ``high-luminosity" QSOs lie in $\sim${\ts}$2L^*$ galaxies (McLeod \&
Rieke 1994b).  H-band imaging of ``cool" + ``warm" ULIGs (typically at $z
< 0.16$), which span a luminosity range sufficient to include all of the
``low-luminosity" QSOs and approximately half of the ``high-luminosity"
QSOs, shows that ULIGs lie in hosts with total H-band luminosities in the
range $\sim${\ts}0.5--5{\ts}$L^*$ with a mean value of $\sim
1.5${\ts}$L^*$ (Surace et al. 1998, 2000).  Data at R-band and K-band for
ULIGs and QSOs is also available, but in the case of ULIGs much of it is
not fully reduced.  However, Figure 7 gives a preliminary comparison of
data for a subsample of ULIGs with ``E-like" profiles with QSOs showing
that both samples have similar R-band and K-band magnitude distributions
and similar R-K colors.

\begin{figure}
\includegraphics{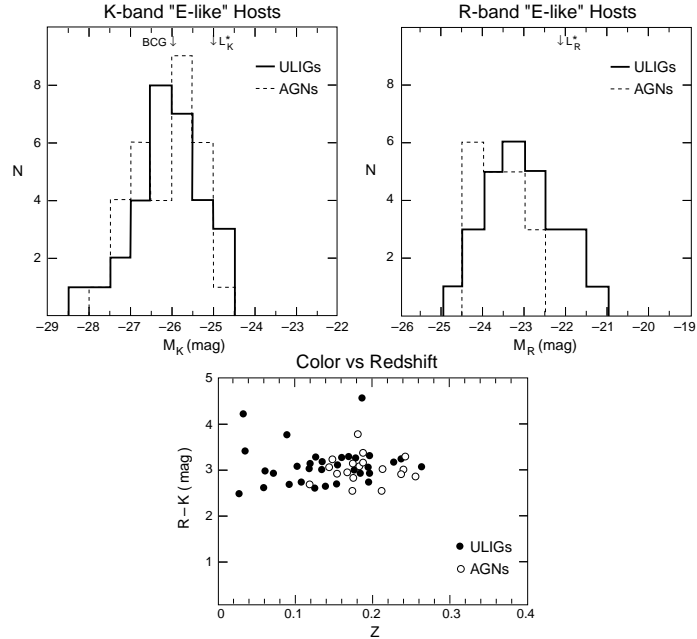}
\vspace{8cm}
\caption{R-band, K-band, and (R$-$K) colors for the hosts of those ULIGs
with single nuclei and $r^{1/4}$-like radial profiles (i.e. ``E-like": see
Kim 1995), compared with recently published K-band and R-band data for the
hosts of AGNs (Taylor et al. 1996; McLure et al. 1999).
} 
\end{figure}

Finally, the radial profiles of QSO hosts have now been published by
several groups (e.g. Hutchings 1987; V{\'e}ron-Cetty \& Woltjer 1990;
McLeod \& Rieke 1994a,b; Taylor et al. 1996; McLure et al. 1999; McLeod \&
McLeod 2001; Percival et al. 2001), with some of the more recent surveys
appearing to solidify many of the claims made in the earlier analyses.
RLQs appear to lie in ``E-like" hosts (as do radio galaxies of similar
radio luminosity).  A significant fraction of RQQs have spiral-like hosts,
but Dunlop (these proceedings) discusses the possibility that this is
simply a luminosity effect, and that above $\sim$3--4{\ts}$L^*$ both RQQs
and RLQs are found almost exclusively in ``E-like" hosts. However,
examination of all of the existing imaging data for QSOs shows that a
significant fraction of all QSOs have radial profiles that are often
somewhat ambiguous, and that even ``E-like" profiles often exclude obvious
tidal debris at large radii as well as observable 2-D structure (e.g.
embedded disks and/or bars) at small radii (e.g. Surace, these
proceedings).

Comparing the 1-D radial profiles of QSOs and ULIGs, it is clear that a
smaller fraction of ULIG hosts are described as ``E-like".  However, given
that the majority of ULIGs are radio-quiet and that the ULIGs studied so
far are heavily biased toward the lower luminosity range of the above QSO
surveys, the results for the two samples may not be all that different.

\section{New data on the Hosts of QSOs}

Here I simply call attention to several important new results, including
those reported at this workshop by our group and others, concerning the
infrared and molecular gas properties and evidence for star formation in
the host galaxies of QSOs.  Evans et al. (2001) and Evans (these
proceedings) have shown that infrared-excess, optically selected QSOs
appear to contain substantial amounts of molecular gas, $M({\rm H}_2) \sim
10^8 - 10^{9.5}{\ts}M_\odot$, similar to the distribution of H$_2$ gas
content found for ``warm" ULIGs.  Surace et al. (2001) and  Surace (these
proceedings) have shown that infrared-excess QSOs exhibit ``knots" of star
formation, and considerable 2D structure in the form of tidal debris
and/or inner bars or disks.  Canalizo \& Stockton (2001) also report that
all ``transition" QSOs (objects which overlap with our samples of ``warm"
ULIGs and infrared-excess QSOs) show evidence for strong recent
star-forming activity and that this activity can be directly related to a
tidal interaction.  In addition, Haas et al. (2000) and Haas (these
proceedings) confirm the existence of ``considerable dust emission" from a
sample PGQSOs that span a wide range of luminosity, and report that the
SEDs of nearly all PGQSOs studied resemble those of ``warm" ULIGs.

\section{Summary and Future Work}

The current data for ULIGs in the local universe provides strong evidence
that the hosts of ULIGs are created by the merger of gas-rich spirals and
that the merger product already is or shortly will be a disky elliptical.
Most of the infrared activity in ULIGs is confined to the inner few kpc
with the larger disk experiencing a decaying starburst.  The debate
continues over the dominant nuclear luminosity source responsible for the
intense infrared emission in the ``cool" ULIGs with some consensus having
been reached that it is an AGN for the ``warm" ULIGs.  In turn, recent
data on lower luminosity RQQs is consistent with their having evolved from
a ``warm" ULIG phase.  However, support for the full evolutionary scenario
(``cool" ULIG$\to$``warm"ULIG$\to$``IR-excess" QSO$\to$QSO at all ULIG/QSO
luminosities and all redshifts remains elusive.

An alternative view of the ULIG$\to$QSO evolutionary scenario is that it
simply does not exist, and that the two populations have essentially
different origins and fates.  Intriguing evidence presented by Dunlop and
collaborators (Dunlop, these proceedings) suggests that the hosts of QSOs
are old ellipticals (e.g. Nolan et al. 2001) where the spheroid (and
presumably the MBH) was put in place long before the current observed QSO
phase (as opposed to a precursor ULIG phase).  The suggestion is then
raised that a minor interaction (or other such fuelling event) has simply
recently re-energised the black hole.  Further observations, perhaps
coupled with better modelling of the merger process that can provide more
accurate age dating of the merger remnant, are clearly needed to resolve
these competing views.



\begin{acknowledgments}

D.B.S. gratefully acknowledges the hospitality of the Max-Plank Institut
f${\rm \ddot u}$r Extraterrestrische Physik and the Alexander
von{\ts}Humboldt Stiftung for a Humboldt senior award.  Partial financial
support was also provided from NASA through JPL Contract no. 961566

\end{acknowledgments}

%


\bibliographystyle{apalike}


\end{document}